\documentclass{article}

\usepackage{arxiv}

\usepackage[utf8]{inputenc} % allow utf-8 input
\usepackage[T1]{fontenc}    % use 8-bit T1 fonts
\usepackage{hyperref}       % hyperlinks
\usepackage{url}            % simple URL typesetting
\usepackage{booktabs}       % professional-quality tables
\usepackage{amsfonts}       % blackboard math symbols
\usepackage{nicefrac}       % compact symbols for 1/2, etc.
\usepackage{microtype}      % microtypography
\usepackage{lipsum}
\usepackage{graphicx}
\graphicspath{ {./images/} }

\title{Universal visibility patterns of unimodal maps}
\author{
 Juan Carlos Nu{\~{n}}o \\
  Department of Applied Mathematics\\ 
    Universidad Polit{\'e}cnica de Madrid\\ 
Madrid - 28040, Spain\\
  \texttt{juancarlos.nuno@upm.es} \\
   \And
 Francisco J. Mu{\~{n}}oz \\
\\
   \texttt{f.j.munoz.ortega@gmail.com}  \\
  \\
 \\
}

\begin{document}
\maketitle
\begin{abstract}
There is order in chaos. This order appears, for instance, at: (i) the fractality of the chaotic attractors, (ii) the universality of the period doubling bifurcations and (iii) the visibility patterns of the time series that are generated by some unimodal maps. In this paper, we focus on the geometric structure of time series that can be deduced from a visibility mapping.  This visibility mapping associates to each point $(t,x(t))$ of the time series to its horizontal visibility basin, i.e. the number of points of the time series that can be seen 
horizontally from each respective point. We apply this study to the class of unimodal maps that give rise to equivalent bifurcation diagrams. We use the classical logistic map to illustrate the main results of this paper: there are visibility patterns in each cascade of the bifurcation diagram, converging  at the onset of chaos.

The visibility pattern of a periodic time series is generated from elementary blocks of visibility in a recursive way. This rule of recurrence applies to all periodic-doubling cascades. Within a particular window, as the growth parameter $r$ varies and each period doubles, these blocks are recurrently embedded forming the visibility pattern for each period. In the limit, at the onset of chaos, an infinite pattern of visibility appears containing all visibility patterns of the periodic time series of the cascade. 

We have seen that these visibility patterns have specific properties: (i) the size of the elementary blocks depends on the period of the time series, (ii) certain time series sharing the same periodicity can have different elementary blocks and, therefore, different visibility patterns, (iii) since the 2-period and 3-period windows are unique, the respective elementary blocks, $ \{2\} $ and $ \{2 \, 3\}$, are also unique and thus, their visibility patterns.  We explore the visibility patterns of other low-periodic time series and also enumerate all elementary blocks of each of their periodic windows. All of these visibility patterns are reflected in the corresponding accumulation points at the onset of chaos, where periodicity is lost.  Finally, we discuss the application of these results in the field of non-linear dynamics.

%\begin{description}
%\end{description}

\end{abstract}

\maketitle

%\tableofcontents

\section{\label{sec:level1}Introduction}

The analysis of time series has become a mature discipline within the field of data analysis. Time series are ubiquos in Nature. Any process that changes over time leaves a fingerprint in the form of a time series. The mathematical modelization of these processes provides equations, either deterministic or stochastic, whose solutions are assumed to fit real realizations. In the deterministic case, these models are written in terms of differential equations (continuous time) or difference equations or maps (discrete time).  They have contributed substantially to the development of the mathematical field of non-linear dynamics, which includes chaos.

It was assumed that simple mathematical models give rise to simple dynamics.  However, the pionner paper of Robert May \cite{May} showed that even simple discrete maps can exhibit very complex dynamics, namely chaotic dynamics.  Many years have passed since the publication some foundational papers like the already cited by May or those published by Feigenbaum \cite{Feigenbaum}, Yorke \cite{Yorke}, Metropoli \cite{Metropoli}, Sharkoskiii \cite{Shark} and many others. From that moment, the study of discrete maps is an active field and still provides interesting mathematical results (see, for instance, \cite{Gilmore}). Besides, this field has also practical relevance for other scientific fields such as data analysis and complexity \cite{Cady}.

The structure behind chaos has been studied since the very beginning of its discovery \cite{Gleick}. Chaotic patterns have attracted the attention of scientists of different fields, from mathematics to physics, as these patterns challenge the general understanding  that dynamics with erratic trajectories could exhibit universal properties that can be described in terms of universal constants. In line with the title of the conference held at Los Alamos in Fall 1982 \cite{LosAlamos}: there is order in chaos. 

The application of statistical physics and complex networks have opened new fruitful lines of research in the time series analysis \cite{Zou}. In particular, the implementation of the visibility algorithm has provided a new geometric viewpoint of the dynamic structure of unimodal maps that exhibit period-doubling bifurcations and chaos \cite{Lacasa1,Luque1,Luque2}. In this context, we have very recently presented a new approach to this analysis that combines the visibility algorithm with set theory and combinatorics to obtain the partial visibility curves of the logistic map \cite{Nuno20}. From this study we have been able to obtain the partial visibility curves at the onset of chaos in the period doubling cascades.  In particular, the length of these curves have been calculated for the 3-cascade and the 2-cascade, the so called, Feigenbaum cascade. The most relevant consequence of these results is the demonstration that the chaotic dynamics hides patterns of visibility that can be obtained as a limit of patterns of visibility of the periodic series in the period doubling cascades. This paper studies precisely the visibility properties of these patterns, putting special emphasis on the description of the elementary blocks that generate the infinite cascades of the bifurcation diagram of unimodal maps.

\begin{figure}[htpb]
\centering
\includegraphics[width=0.8\columnwidth]{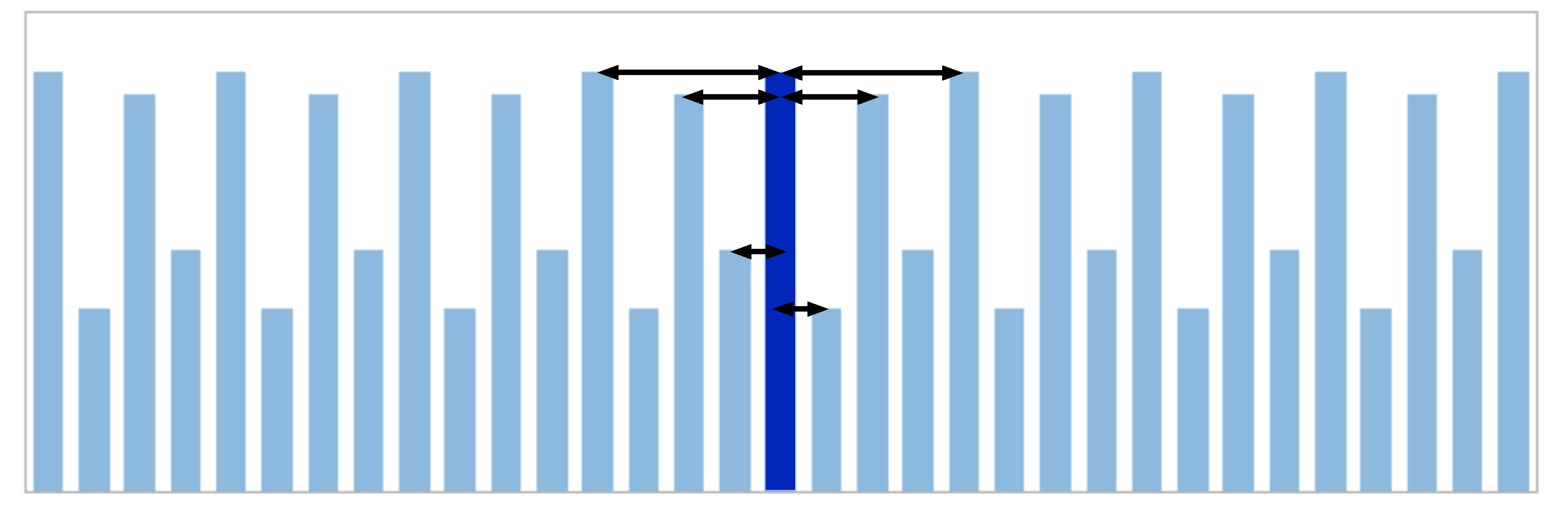}
\caption{Schematic representation of the horizontal visibility map. The middle point of the time series sees horizontally 6 other points that are connected by arrows. The visibility map associates to each point of the time series the cardinal of its basin of visibility. Note that the period of the time series is $T=4$ and that the visibility pattern $\{6 \, 2 \, 4 \, 2 \}$ is repeated.}
\label{figv1}
\end{figure}

A time series can be generated as a solution of an Initial Value Problem associated to the  following difference equation :
\begin{equation}
x_{n} = f(x_{n-1}; \mu)
\end{equation}
for $n = 1,2, \ldots$ with the initial condition $x_0$. This map generates an infinite sequence of real numbers $\{x_n\}_{n \in \mathbb{N}}$ that, if $n$ represents time, is usually referred as time series. 
It is worth remembering that the time series forms an ordered set which is different from the set of points that is generated from the map.  In this work we will use the so called {\it horizontal visibility} \cite{Luque3}: Two points of the time series  $(t_i,x_i)$  and $(t_j, x_j)$ such that $t_i < t_j$ see horizontally each other if $x_k  < min\{x_i,x_j\}$ for all $t_k \in (t_i,t_j)$ (see Fig. \ref{figv1}).  In addition, we use the following concepts: The {\it basin of visibility of a point} is the set of points seen from a point $P_k(k,x_k)$ of the time series contour. The cardinality of the basin of visibility of a point $P_k(k,x_k)$ is its {\it visibility}, $v_k$.   The {\it visibility distribution of the time series} is the distribution function of the visibility of all of the points that form the times series: for each visibility number, $v$, the distribution $P(v)$ represents the number of points of the time series with that visibility. 

\begin{figure}
\centering
\includegraphics[width=0.8\columnwidth]{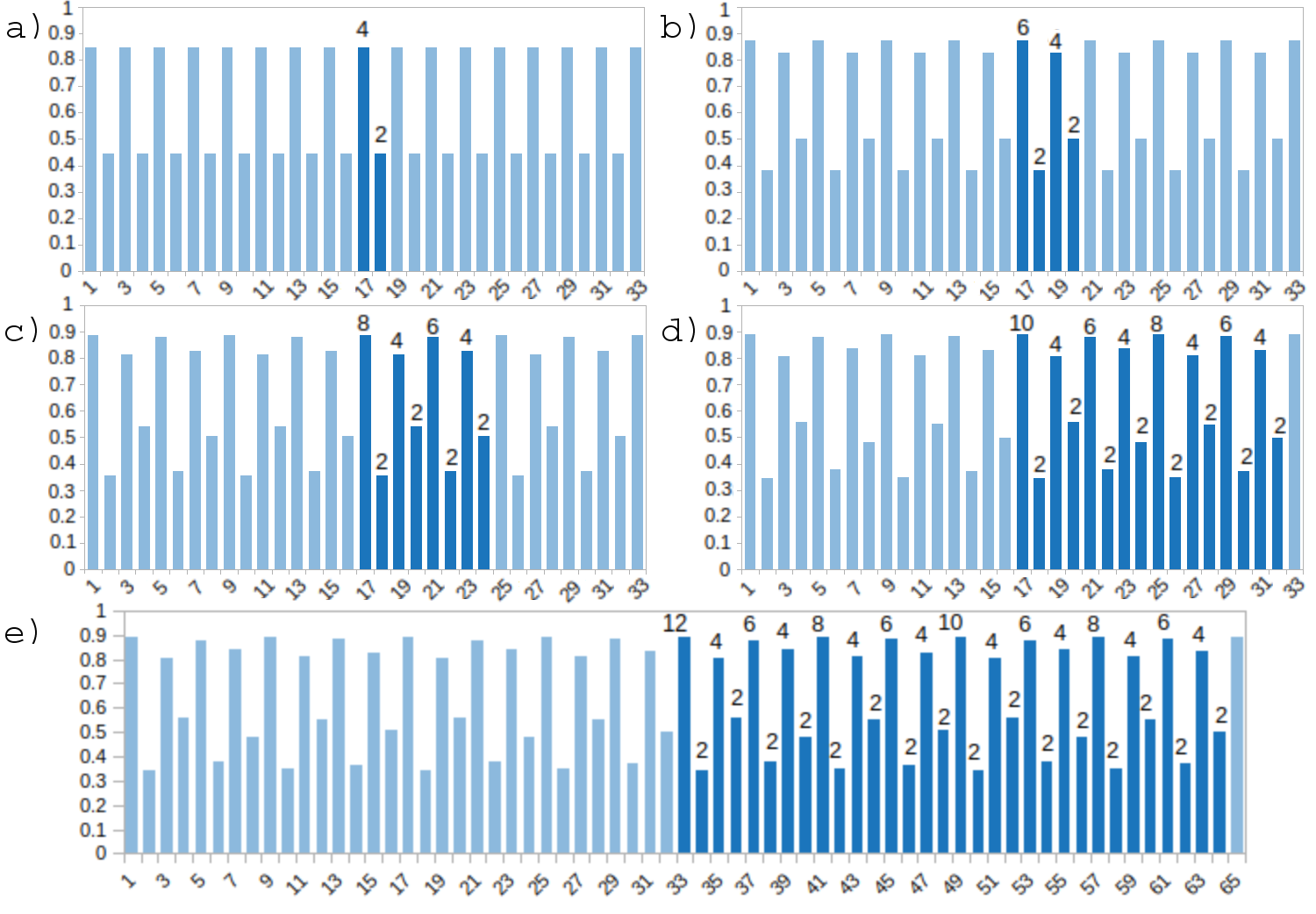}
\caption{Five time series with different periodicities: (a) $T=2$,  (b) $T=2^2$, (c) $T=2^3$, (d) $T=2^4$ and (e) $T=2^5$, within the Feigenbaum cascade where the visibility patterns are highlighted. Note that the visibility pattern of the bottom row time series can be obtained from the previous visibility patterns. The time series are generated from the logistic map with (a) $r=3.2$, (b)  $r=3.5$,  (c) $r=3.55$, (d) $r=3.566$ and (e) $r=3.5690$.}
\label{Fig1}
\end{figure}

Unimodal maps are a particular class of maps generated by a one-dimensional continuous function $f: I=[0,1] \to [0,1]$ with a unique local maximum in $c \in (0,1)$, thus  $f[0,c)$ is strictly increasing and $f(c,1]$ is strictly decreasing. In addition, it is usually assumed that $f(0) =f(1) = 0$. This kind of maps have been widely studied since they exhibit chaotic behavior (see, for example, \cite{Schuster, Strogatz, Peitgen}).  Three well known examples of unimodal maps:  (i) The quadratic or {\it logistic} map $f(x;r) = r \, x \, (1 - x)$,  (ii) the {\it Sine} map: $f(x;r) = r \, sin (\pi \, x)$ and (iii)  the {\it Tent} map, defined as a piecewise function: $f(x) = 2\,r \, x$ for $0 \leq x \leq \frac{1}{2}$ and $f(x;r) = 2\,r \, (1 - x)$ for $\frac{1}{2} < x \leq 1$ for $r\in[0,1]$. This latter map shares most of the properties of unimodal maps, despite its singularity. Note that with these definitions the maxima of the tree maps take the same value and are attained at the same value, i.e. $f_{max}(1/2)=1$.  In all of these maps, as the parameter $r$ varies two types of bifurcations take place systematically: saddle–node and period-doubling bifurcations. Periodic orbits are formed in saddle–node bifurcations. It turns out that at some point the stable node loses its stability giving rise to a period-doubling cascade \cite{Tucker}. 

The universality of the structural properties of these maps were first established by Feigenbaum in 1978 \cite{Feigenbaum}. Taking the value of the so called Feigenbaum indices, these one-dimensional maps can be classified into different universality classes \cite{Livadiotis}. In practice, sharing the same universality class means that there is a one-to-one correspondence between conjugate maps and that these maps share the same behavior \cite{Gilmore}.  The bifurcation diagram of unimodal maps exhibit an infinite number of periodic windows. Several procedures have been proposed to find these windows \cite{Galias, Livadiotis}. The number of periodic windows for each period $T$ is given in table II of reference \cite{Galias}. As it can be seen, the number of windows increases sharply with the period $T$. For ease of convenience, we focus on lower period cascades to show the visibility patterns and their main properties. Besides, in most cases, we will use the logistic map to illustrate the main results.

\begin{figure} [htpb]
\centering
\includegraphics[width=1.0\columnwidth]{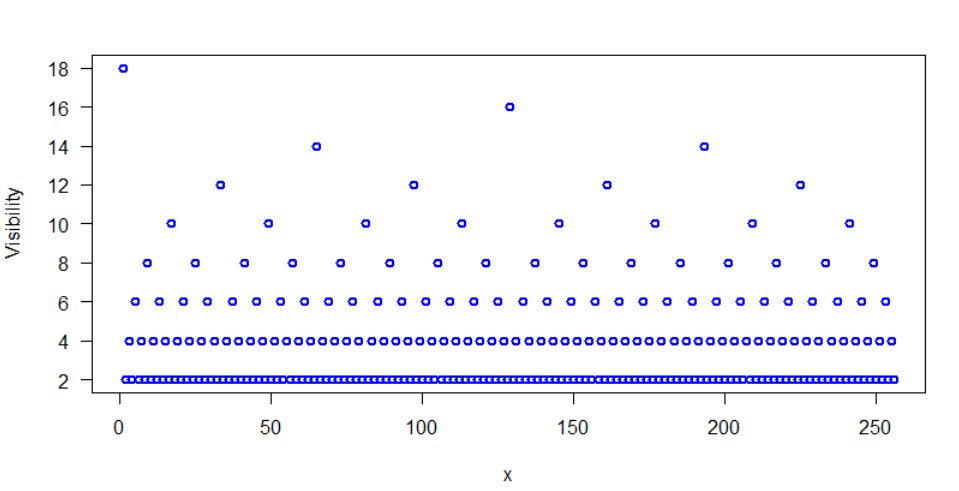}
\caption{Visibility pattern for a period $T=2^8$ time series in the Feigenbaum cascade. These points are generated by a recurrent procedure that enables the generation of the visibility pattern of any period. In this case, the largest visibility is $V_{max} = 18 = 2\, (8 + 1)$.  The number of points in the pattern is 256. For a time series of size $N$, this pattern is repeated approximately $N\,2^{-8}$ times.}
\label{Fig2}
\end{figure}

\section{\label{sec4:level1}Visibility pattern at the onset of chaos of the Feigenbaum cascade}

As mentioned above, in all unimodal maps described in the previous section a doubling period bifurcation appears as the growth rate $r$ increases. For instance, in the logistic map, this period-doubling cascade starts at $r=3$ and ends at the onset of chaos at the so called Feigenbaum accumulation point $r_{\infty} = 3.5699457...$.  In figure \ref{Fig1} we show the visibility patterns of certain low period time series and highlight the recurrent procedure to obtain the pattern of period $T=2^{n+1}$ from the period $T=2^n$. The elementary block for the Feigenbaum cascade is $C_2=\{2\}$. This block of size $n=1$ is embedded in the visibility pattern of the next non-elementary block of size $n=2$, after the largest visibility, which in this case is 4. The procedure is repeated successively $n> 2$ (see table \ref{tab1}). For each periodic time series, this visibility pattern is repeated infinitely or, in practice, while the finite size of the time series allows it.   This recursive structure of elementary blocks is clearly reflected in Figure \ref{Fig2} for the period $T=2^8$.  For convenience, all blocks start by the largest visibility (basin), $v_{max}$. This largest visibility basin increases linearly with $n$, specifically as $v_{max}(n) =2(n+1)$ for $n=1,2,\ldots$. At the limit, the onset of chaos, the time series contains all of the patterns of the periodic series forming a non-periodic series whose largest visibility (basin) tends to infinity. Needless to say that, in practice, this infinite pattern would only be perceived if the time series has infinite size.

%\begin{widetext}
\begin{table}[tpb]
\label{tab1}
\caption{Visibility patterns for the period doubling Feigenbaum cascade. A recurrent procedure  is applied for the construction of the visibility patterns for the successive periods $T=2^{n}$. As it can be seen, the next visibility pattern contains the previous one after the elementary block is inserted. In the limit, the visibility pattern at the onset of chaos would contain the visibility patterns of all previous periods. } 
 \small
\begin{tabular}{ll}
\hline  
\textbf{n} & \textbf{Pattern}    \\
\textbf{0} & 2   \\

\textbf{1} & 4 2   \\

\textbf{2} & 6 2 4 2          \\

\textbf{3} & 8 2 4 2 6 2 4 2 \\

\textbf{4} & 10 2 4 2 6 2 4 2 8 2 4 2 6 2 4 2 \\

\textbf{5} & 12 2 4 2 6 2 4 2 8 2 4 2 6 2 4 2 10 2 4 2 6 2 4 2 8 2 4 2 6 2 4 2 \\

\ldots          &     \\

\textbf{n} &   (2 n+2) 2 4 2 6 2 4 2 8 2 4 2 6 2 4 2 10 2 4 2 6 2 4 2  \ldots (2 n) 2 4 6 2 4 2 8 2 4 2 6 2 4 2 ...  \\

\hline 
\end{tabular}
\end{table}
%\end{widetext}

\begin{figure} 
\centering
\includegraphics[width=0.8\columnwidth]{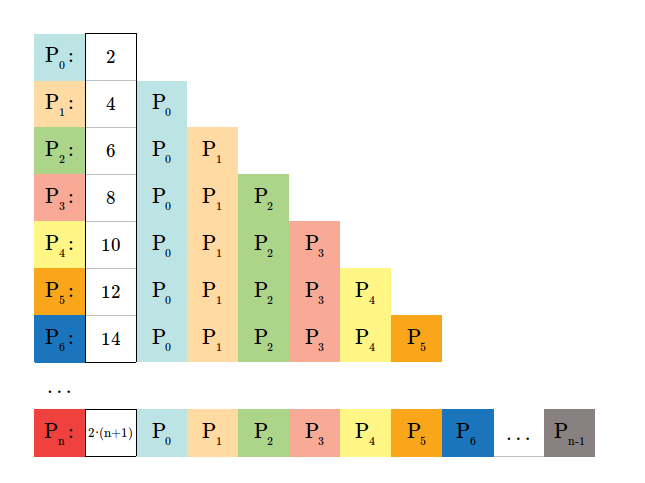}
\caption{Schematic representation of the procedure for the generation of visibility patterns in the period doubling Feigenbaum cascade. A recurrencce formula enables the calculation of a visibility pattern from the previous ones; the elementary block is always inserted after the largest visibility, followe by the rest of the previous patterns. At the limit  $n \to \infty$ we would obtain the visibility pattern of this cascade at the onset of chaos of this cascades, $P_{\infty}$, for any unimodal map.}
\label{Fig3}
\end{figure}

A recursive law for the formation of the visibility pattern of the period $T=2^n$ is straightforwardly derived (see Figure \ref{Fig3}). If $P_{n-1}$ is the visibility pattern for $n=1,2,\ldots$, then the visibility pattern for the next period is given by:

\[
P_n = 2(n+1) \{2\} P_1 P_2 P_3 \ldots P_{n-1}  \ldots
\]

Notice the recurrence law that generates the $nth$ visibility pattern $P_n$: The largest visibility precedes the elementary block, $\{ 2 \}$ and after, all previous visibility patterns follow until the one  before, $P_{n-1}$, that contains again all previous visibility patterns. In a certain sense, it reminds the Matryoshka dolls formed by a (finite) sequence of nesting dolls.

\section{\label{sec4:level1}Visibility pattern at the onset of chaos of the period-3 cascade}

In addition to the Feigenbaum cascade, unimodal maps exhibit a unique 3-period window where  time series of periods $T=3 \, 2^{n-1}$ appears. As explained above, we use the symbolic dynamics induced by the visibility algorithm to determine the visibility patterns of the periodic time series and at the onset of chaos. It can be shown that, in this cascade,  the elementary block is $C_3=\{2 \, 3\}$. Similarly to the Feigenbaum cascade, this $nth$ visibility pattern is recursive and it can be described in terms of the pevious pattern as:
\[
P_{n} = (2(n+1)+1) \{2 \, 3\} P_1 P_2 P_3 P_4 \ldots P_{n-1} \ldots
\]
being $P_k$ defined in table \ref{tab2}.

%\begin{widetext}
\begin{table}[tpb]
\caption{Visibility patterns for 3-period doubling cascade. As in the previous table, a recurrent procedure is applied for the construction of the visibility patterns for the successive periods $T=3\,2^{n-1}$ for $n=1,2,\ldots$.  The $nth$-visibility pattern $P_n$ is formed after inserting the elementary block $\{2 \, 3 \}$ and all previous visibility patterns. At the limit, the visibility pattern at the onset of chaos of this cascade would contain the visibility patterns of the infinite lower periods. } 
\label{tab2}
 \small
\begin{tabular}{ll}
\hline  
\textbf{n} & \textbf{Pattern}    \\

% \textbf{0} & 2 3 \\

\textbf{1} & 5 2 3    \\

\textbf{2} & 7 2 3 5 2 3         \\

\textbf{3} & 9 2 3 5 2 3 7 2 3 5 2 3   \\

\textbf{4} & 11 2 3 5 2 3 7 2 3 5 2 3 9 2 3 5 2 3 7 2 3 5 2 3 \\

\textbf{5} & 13 2 3 5 2 3 7 2 3 5 2 3 9 2 3 5 2 3 7 2 3 5 2 3 11 2 3 5 2 3 7 2 3 5 2 3 9 2 3 5 2 3 7 2 3 5 2 3 \\

\ldots          &     \\

\textbf{n} &   (2n+3) 2  3  5  2  3  7  2  3  5  2  3  9  2  3  5  2  3  7  2  3  5  2  3  \ldots (2n+1) 2 3 5 2 3 7 2 3 5 2 3 \ldots \\
\hline 
\end{tabular}
\end{table}
%\end{widetext}

\section{\label{sec4:level1}Visibility patterns of other low period-doubling cascades}

Visibility patterns can be obtained for all periodic windows of the bifurcation diagram of unimodal maps. The few limitations in practice relate to the computational efforts requiered to perform the calculations. In this section, we describe the visibility patterns of time series belonging to low period windows. For $T=4$, two periodic windows exist: (i) one within the Feigenbaum cascade and  (ii) the other one formed by period doublings of period 4, i.e. $T = 4 \, 2^{n-1}$ for $n=1,2,\ldots$. The elementary block corresponding to this 4-period cascade is $C_4^2 =\{2  \, 3 \, 3 \}$ and, as mentioned, the successive visibility patterns of the cascade is formed by recurrence as depicted in table \ref{tab3}. Note that, as in the previous cascades, the largest visibility for each period is $2 (n+1) + 2 $.  

%\begin{widetext}
\begin{table}[tpb]
\caption{As in the previous tables, we show the visibility patterns for 4-period doubling cascade: $T=4\,2^{n-1}$ for $n=1,2,\ldots$. The procedure of generation of the successive patterns is the same as the one applied to other cascades, but it starts from a different elementary block: $\{ 2 \, 3 \, 3\}$. At the limit of infinite periodicity, we would obtain the visibility pattern of the onset of chaos of this cascade} 
\label{tab3}
 \small
\begin{tabular}{ll}
\hline  

\textbf{n} & \textbf{Pattern}    \\

% \textbf{0} & 2 3 3 \\

\textbf{1} & 6 2 3 3    \\

\textbf{2} & 8 2 3 3 6 2 3 3         \\

\textbf{3} & 10 2 3 3 6 2 3 3  8 2 3 3 6 2 3 3  \\

\textbf{4} & 12 2 3 3  6 2 3 3 8 2 3 3 6 2 3 3 10 2 3 3 6 2 3 3  8 2 3 3 6 2 3 3 \\

\textbf{5} & 14 2 3 3 6 2 3 3 8 2 3 3 6 2 3 3 10 2 3 3 6 2 3 3  8 2 3 3 6 2 3 3 12 2 3 3   \\
& 6 2 3 3 8 2 3 3 6 2 3 3 10 2 3 3 6 2 3 3  8 2 3 3 6 2 3 3  \\

\ldots          &     \\

\textbf{n} &   (2n+4) 2 3 3 6 2 3 3 8 2 3 3 6 2 3 3 10 2 3 3 6 2 3 3  8 2 3 3 6 2 3 3  \ldots (2n+2) 2 3 3 6 2 3 3 8 2 3 3 
\ldots \\

\hline 
\end{tabular}
\end{table}
%\end{widetext}

There are three fundamental 5-period cascades with periods: $T=5\,2^{n-1}$. These windows are separated in three intervals of the bifurcation diagram. Their exact location depend on the particular unimodal map.  
Only two elementary blocks exist for these cascades:
\[ 
C_5^1 : \{2 \, 3  \,4 \,2 \} ; \, C_5^2 :  \{2 \, 3  \, 3 \, 3 \}
\]
For instance, in the logistic map, the elementary block $C_5^1$ generates the visibility patterns in the $r$-intervals 
$ [3.738173...,3.74112... ]$ and $ [3.905572...,3.906107... ]$. The other $C_5^2$ gives rise to the visibility patterns of the 5-period doubling cascade that occurs in the interval $  [3.990258...,3.990296... ]$ (see table \ref{tab7})

The recursive procedure that theoretically generates the visibility pattern for each period within each of the cascades is similar to the one previously described (table \ref{tab4}), that is adding each elementary block to the largest visibility for each  $n=1,2,\ldots$, followed by the previous visibility patterns until reaching the one before.  Following this approach allows us to obtain the recurrent formula for the $n$-visibility pattern, $P_n$:
\[
P_n= (2 (n+1) + 3) C_5 P_1 P_2 P_3 P_4 \ldots P_{n-1} \ldots
\]
where $C_5$ can be either one of the two elementary blocks corresponding to these 5-period cascades. A similar table can be drawn for the visibility patterns deriving from the elementary block $C_5^2$.  

%\begin{widetext}
\begin{table}[tpb]
\caption{Visibility patterns for 5-period doubling cascade: $T=5\,2^{n-1}$ for $n=1,2, \ldots$, derived from the elementary block $C_5^1=\{ 2 \, 3  \,4 \,2 \}$. A similar table could be obtained for the other elementary block of basic period $m=5$, i.e. $C_5^2$.} 
\label{tab4}
 \small
\begin{tabular}{ll}
\hline  

\textbf{n} & \textbf{Pattern}    \\

% \textbf{0} & 2 3 4 2 \\

\textbf{1} & 7 2 3 4 2    \\

\textbf{2} & 9 2 3 4 2 7 2 3 4 2     \\

\textbf{3} & 11 2 3 4 2 7 2 3 4 2 9 2 3 4 2 7 2 3 4 2   \\

\textbf{4} & 13 2 3 4 2 7 2 3 4 2 9 2 3 4 2 7 2 3 4 2 11 2 3 4 2 7 2 3 4 2 9 2 3 4 2 7 2 3 4 2 \\

\textbf{5} & 15 2 3 4 2 7 2 3 4 2 9 2 3 4 2 7 2 3 4 2 11 2 3 4 2 7 2 3 4 2 9 2 3 4 2 7 2 3 4 2 \\
& 13 2 3 4 2 7 2 3 4 2 9 2 3 4 2 7 2 3 4 2 11 2 3 4 2 7 2 3 4 2 9 2 3 4 2 7 2 3 4 2  \\

\ldots          &     \\

\textbf{n} &  (2n+5) 2 3 4 2 7 2 3 4 2 9 2 3 4 2 7 2 3 4 2 11 2 3 4 2 7 2 3 4 2 9 2 3 4 2 7 2 3 4 2  \ldots \\
& (2n+3) 2 3 4 2 7 2 3 4 2 9 2 3 4 2 7 2 3 4 2  \ldots \\

\hline 
\end{tabular}
\end{table}
%\end{widetext}

We now consider basic period $T=6$. There are five windows in the bifurcation diagram of unimodal maps where the time series have this period. One of these intervals falls inside the 3-period cascade. The other four are fundamental cascades that starts with the basic period 6. There are only three elementary blocks that generate the visibility patterns of the four period doubling cascades:
$$
C_6^1 = \{ 2 \, 4 \,2 \,5 \, 2 \}; \, C_6^2 =  \{ 2 \, 3 \, 3 \, 4 \, 2  \}; \, C_6^3  =  \{2 \, 3 \, 3 \, 3 \, 3 \}
$$ 
In contrast to previous basic periods, the basic visibility patterns for period $T=6$ have different maximum visibilities (see tables \ref{tab61},\ref{tab62},\ref{tab63}). One of them, generated from $C_6^1$, similar to the visibility pattern for the periodic series within the 3-period cascade, has a maximum visibility of 7, whereas the other two have a maximum visibility of 8 (i.e. $6 + 2$). In any case, as mentioned, the formula of recurrence for the generation of visibility patterns for any period  $n>2$ within the cascades coincides with the previous periods: after the largest visibility, the elementary block is inserted and then, it is followed by the previous visibility patterns until $P_{n-1}$, which, in turn, contains all the previous visibility patterns.

%\begin{widetext}
\begin{table}[tbp]
\caption{Visibility patterns for 6-period doubling cascade, $T=6\,2^{n-1}$, generated from the elementary block $C_6^1=\{ 2 \, 4 \,2 \,5 \, 2 \}$.  Note that, contrarily to the previous cases $m=2,3,4,5$, the largest visibility is not given by the formula: $2 (n + m)$. This is due to the constraint imposed by the total visibility of the elementary block \ref{TotV} that for this elementary block yields $V_{max} = 7$ for $P_1$. Applying the recursive procedure to this pattern we obtain that $V_{max}$ for the $nth$-pattern is $2 (n + 5)$. }
\label{tab61}
 \small
\begin{tabular}{ll}
\hline  

\textbf{n} & \textbf{Pattern}    \\

% \textbf{0} & 2 4 2 5 2  \\

\textbf{1} & 7 2 4 2 5 2    \\

\textbf{2} & 9 2 4 2 5 2 7 2 4 2 5 2     \\

\textbf{3} & 11 2 4 2 5 2 7 2 4 2 5 2 9 2 4 2 5 2 7 2 4 2 5 2    \\

\textbf{4} & 13 2 4 2 5 2 7 2 4 2 5 2 9 2 4 2 5 2 7 2 4 2 5 2 11 2 4 2 5 2 7 2 4 2 5 2 9 2 4 2 5 2 7 2 4 2 5 2  \\

\textbf{5} & 15 2 4 2 5 2 7 2 4 2 5 2 9 2 4 2 5 2 7 2 4 2 5 2 11 2 4 2 5 2 7 2 4 2 5 2 9 2 4 2 5 2 7 2 4 2 5 2 \\
& 13 2 4 2 5 2 7 2 4 2 5 2 9 2 4 2 5 2 7 2 4 2 5 2 11 2 4 2 5 2 7 2 4 2 5 2 9 2 4 2 5 2 7 2 4 2 5 2 \\

\ldots          &     \\

\textbf{n} &  (2n+5) 2 4 2 5 2 7 2 4 2 5 2 9 2 4 2 5 2 7 2 4 2 5 2 11 2 4 2 5 2 7 2 4 2 5 2 9 2 4 2 5 2 7 2 4 2 5 2 \\
& 13 2 4 2 5 2 7 2 4 2 5 2 9 2 4 2 5 2 7 2 4 2 5 2 11 2 4 2 5 2 7 2 4 2 5 2 9 2 4 2 5 2 7 2 4 2 5 2 \ldots \\
& (2n+3) 2 3 4 2 7 2 3 4 2 9 2 3 4 2 7 2 3 4 2  \ldots \\

\hline 
\end{tabular}
\end{table}
%\end{widetext}

%\begin{widetext}
\begin{table}[tbp]
\caption{Visibility patterns for 6-period doubling cascade: $T=6\,2^{n-1}$, generated from the elementary block $C_6^2= \{ 2 \, 3 \, 3 \, 4 \, 2  \}$. Contrary to the visibility patterns generated from $C_6^1$ (see table \ref{tab61}), the largest visibility follows the formula $2\,(n+m)$, where $m=6$.  The procedure for the generation of the $nth$-visibility pattern is universal for any basic period $m$.} 
\label{tab62}
 \small
\begin{tabular}{ll}
\hline  

\textbf{n} & \textbf{Pattern}    \\

% \textbf{0} & 2 3 3 4 2  \\

\textbf{1} & 8 2 3 3 4 2    \\

\textbf{2} & 10 2 3 3 4 2 8 2 3 3 4 2     \\

\textbf{3} & 12 2 3 3 4 2 8 2 3 3 4 2 10 2 3 3 4 2 8 2 3 3 4 2    \\

\textbf{4} & 14 2 3 3 4 2 8 2 3 3 4 2 10 2 3 3 4 2 8 2 3 3 4 2 12 2 3 3 4 2 8 2 3 3 4 2 10 2 3 3 4 2 8 2 3 3 4 2   \\

\textbf{5} & 16 2 3 3 4 2 8 2 3 3 4 2 10 2 3 3 4 2 8 2 3 3 4 2 12 2 3 3 4 2 8 2 3 3 4 2 10 2 3 3 4 2 8 2 3 3 4 2 \\
& 14 2 3 3 4 2 8 2 3 3 4 2 10 2 3 3 4 2 8 2 3 3 4 2 12 2 3 3 4 2 8 2 3 3 4 2 10 2 3 3 4 2 8 2 3 3 4 2  \\

\ldots          &     \\

\textbf{n} &  (2n+6) 2 3 3 4 2 8 2 3 3 4 2 10 2 3 3 4 2 8 2 3 3 4 2 12 2 3 3 4 2 8 2 3 3 4 2 10 2 3 3 4 2 8 2 3 3 4 2 \\
& 14 2 3 3 4 2 8 2 3 3 4 2 10 2 3 3 4 2 8 2 3 3 4 2 12 2 3 3 4 2 8 2 3 3 4 2 10 2 3 3 4 2 8 2 3 3 4 2  \ldots \\
& (2n+4)  2 3 3 4 2 8 2 3 3 4 2 10 2 3 3 4 2 8 2 3 3 4 2 \ldots \\

\hline 
\end{tabular}
\end{table}
%\end{widetext}

%\begin{widetext}
\begin{table}[tbp]
\caption{Visibility patterns for 6-period-doubling cascade, $T=6\,2^{n-1}$, generated from the elementary block $C_6^3=\{2 \, 3 \, 3 \, 3 \, 3 \}$. As in table \ref{tab62}, the largest visibility for the $nth$-pattern is given by $2\, (n+6)$, according with the visibility constraint of the elementary blocks given by formula \ref{TotV}. Note that, as stated before, when $n \to \infty$ we would obtain the visibility pattern at the onset of chaos  that would contain the infinite visibility patterns of all the periods of this cascade.} 
\label{tab63}
 \small
\begin{tabular}{ll}
\hline  

\textbf{n} & \textbf{Pattern}    \\

% \textbf{0} &  2 3 3 3 3  \\

\textbf{1} & 8 2 3 3 3 3    \\

\textbf{2} & 10 2 3 3 3 3 8 2 3 3 3 3      \\

\textbf{3} & 12 2 3 3 3 3 8 2 3 3 3 3 10 2 3 3 3 3 8 2 3 3 3 3    \\

\textbf{4} & 14 2 3 3 3 3 8 2 3 3 3 3 10 2 3 3 3 3 8 2 3 3 3 3 12 2 3 3 3 3 8 2 3 3 3 3 10 2 3 3 3 3 8 2 3 3 3 3   \\

\textbf{5} & 16 2 3 3 3 3 8 2 3 3 3 3 10 2 3 3 3 3 8 2 3 3 3 3 12 2 3 3 3 3 8 2 3 3 3 3 10 2 3 3 3 3 8 2 3 3 3 3 \\

& 14 2 3 3 3 3 8 2 3 3 3 3 10 2 3 3 3 3 8 2 3 3 3 3 12 2 3 3 3 3 8 2 3 3 3 3 10 2 3 3 3 3 8 2 3 3 3 3   \\

\ldots          &     \\

\textbf{n} &  (2n+6) 2 3 3 3 3 8 2 3 3 3 3 10 2 3 3 3 3 8 2 3 3 3 3 12 2 3 3 3 3 8 2 3 3 3 3 10 2 3 3 3 3 8 2 3 3 3 3 \\
& 14 2 3 3 3 3 8 2 3 3 3 3 10 2 3 3 3 3 8 2 3 3 3 3 12 2 3 3 3 3 8 2 3 3 3 3 10 2 3 3 3 3 8 2 3 3 3 3  \ldots \\
& (2n+4) 2 3 3 3 3 8 2 3 3 3 3 10 2 3 3 3 3 8 2 3 3 3 3 \ldots \\

\hline 
\end{tabular}
\end{table}
%\end{widetext}

There are 9 windows of period 7. All of the periodic orbits in each window are generated from the basic period 7, by period doubling $T=7\,2^{n-1}$ for $n=1,2,\ldots$. These windows are located at different intervals of the bifurcation diagram. Table \ref{tab7} shows these $r$-intervals for the logisitic map. There are only four elementary blocks that generate the visibility patterns in these period 7 windows:
\[
C_7^1= \{2 \, 3 \, 4 \, 2 \, 5 \, 2 \}; \hspace*{5mm} C_7^2= \{2 \, 3 \, 5 \, 2 \, 4 \, 2 \}; \hspace*{5mm} C_7^3= \{2 \, 3 \, 3 \, 5 \, 2 \, 3 \}; \hspace*{5mm} C_7^4= \{2 \, 3 \, 3 \, 3 \, 4 \, 2 \}
\] 

All visibility patterns for the periodic orbits of this period doubling cascade can be generated  from these elementary blocks.  The largest visibility of the visibility pattern corresponding to the basic period 7 are:  $8$ for $C_7^1$, $C_7^2$, and $C_7^3$ and 9 for $C_7^4$ (see table \ref{tab7}). For example, the visibility pattern, $P_1= 8 2 3 4 2 5 2 $, of any time series of period 7 obtained from the elemetary block $C_7^1$ appears in two $r$-intervals of the logistic map: $ [3.701641...,3.702154... ]$ and $ [3.922186...,3.922215... ]$.

We end the presentation of the visibility patterns with the 8-periodic windows. There are 16 periodic windows of period 8. Two of them derive from the period doubling cascades of the basic periods  $m=2$ ($T=2^3$) and $m=4$ ($T=4 \, 2 = 8$) and, hence, the corresponding visibility patterns are directly obtained from the elementary blocks $C_2= \{2 \}$ and $C_4^2 =  \{2 \, 3 \, 3 \}$. More specifically, the visibility patterns of these basic periods are: $P_1^1= \{8 \, 2 \, 4 \, 2 \, 6 \, 2 \, 4 \, 2 \}$ and $P_1^2= \{8 \, 2 \, 3 \, 3 \, 6 \, 2 \, 3 \, 3 \}$ and they appear in the $r$-intervals of the logistic maps given in table \ref{tab7}. The remaining elementary blocks and the corresponding $r$ intervals for the logistic map are shown in table \ref{tab7}. In total, there are 7 distinct elementary blocks:
$C_8^3 =  \{2 \, 4 \, 2 \, 5  \, 2 \, 5 \, 2 \}$; $C_8^4 =  \{2 \, 3 \, 6 \, 2 \, 3 \, 4 \, 2 \}$; $C_8^5 =  \{2 \, 3 \, 4 \, 2 \, 6 \, 2 \, 3 \}$; $C_8^6= \{2 \, 3 \, 3 \, 4 \, 2 \, 5 \, 2 \}$; $C_8^7= \{2 \, 3 \, 3 \, 5 \, 2 \, 4 \, 2 \}$; $C_8^8= \{2 \,3 \, 3 \, 6  \,2  \,3  \,3 \}$; $C_8^9= \{2 \, 3  \,3  \, 3  \,5  \,2  \, 3 \}$; $C_8^{10}= \{2 \, 3  \, 3  \, 3 \, 3 \, 4 \, 2 \}$.

Before ending this section, it might be worth mentioning that the sum of the visibilities of the points of the visibility pattern for the basic period of the cascade,  $V_T$, follows a linear dependence with the period $T$:
\begin{equation}\label{TotV}
V_T = 4 \, T - 2
\end{equation}
This means that, for example, all of the visibility patterns of period $T=6$ have a total visibility $V_T = 22$. If the sum of the visibility of the elementary bolcks is $V_P$, then the largest visibility in the visibility pattern of the first period of the cascade, $V_{max}$,  is given by:
\[
V_{max} =V_T - V_P
\]

Consequentely, we can generate the visibility patterns of each period-doubling cascade of basic period $m$ ($T=m\,2^{n-1}$ for $n=1,2,\ldots$)  by simply knowing its elementary block.  If the elementary block has $m-1$ elements, it corresponds to a basic period $m$. For each $m$ and $n$ we know the total visibility of the elementary blocks $V_P$ and the total visibility of the basic period visibility pattern $V_T$. We can therefore compute $V_{max}$ using the previous expression. As the period increases, we know that the largest visibility increases linearly, $V_{max}^n = V_{max} + 2\,n$ for $n=1,2,\ldots$.
For example,   we can calculate the largest visibility element of the visibility pattern of the basic period $T=8$ since we know that the total visibility of the visibility pattern is $V_T=30$,. Thus, from the elementary block $C_8^9$ we obtain $V_{max} =9$ and the corresponding visibility pattern of the basic period: $P_1^9=\{ 9 \, 2 \, 3 \, 3 \, 3 \, 5 \,  2 \, 3 \}$. The successive visibility patterns for the period doubling time series can be obtained from the recursive formula as shown above.

\begin{table}[tbp]
\caption{Visibility patterns of the basic periods $m=1,2,3,\ldots,8$ and the $r$-intervals of appearance in the bifurcation diagram of the logisitic map. As obtained numerically, the interval bounds are approximate. Note that in all the cases, the formula \ref{TotV} applies.  This explains that visibility patterns with different maximum visibility appear for the same basic period $m$. We could extent this table to capture larger basic periods, but the number of periodic windows increases drastically with $m$ \cite{Galias}.}
\label{tab7}
\begin{tabular}{llll}  
\hline
\textbf{Period} & \textbf{Left bound}   & \textbf{Right  bound}          & \textbf{Visibility Pattern}                  \\
\hline
1       & 1           & 3                                & 2                       \\
2       & 3           & 3.44949                & 4 2                    \\
4       & 3.44949     & 3.544090359            & 6 2 4 2              \\
8       & 3.544090359 & 3.564406325           & 8 2 4 2 6 2 4 2  \\
6       & 3.626554    & 3.630388                   & 7 2 4 2 5 2        \\
8       & 3.662108925 & 3.66244065           & 8 2 4 2 5 2 5 2  \\
7       & 3.701641    & 3.702154                 & 8 2 3 4 2 5 2     \\
5       & 3.738173    & 3.74112                    & 7 2 3 4 2           \\
7       & 3.774134    & 3.774455                  & 8 2 3 5 2 4 2     \\
8       & 3.800740125 & 3.800863725           & 8 2 3 6 2 3 4 2  \\
3       & 3.828428    & 3.841498                   & 5 2 3                 \\
6       & 3.8415      & 3.84761                     & 7 2 3 5 2 3        \\
8       & 3.870532125 & 3.87056755    & 8 2 3 6 2 3 4 2  \\
7       & 3.886029    & 3.886097        & 8 2 3 5 2 4 2     \\
8       & 3.899462425 & 3.899488625  & 8 2 3 4 2 6 2 3  \\
5       & 3.905572    & 3.906107                  & 7 2 3 4 2           \\
8       & 3.912042025 & 3.912060375   & 8 2 3 4 2 6 2 3  \\
7       & 3.922186    & 3.922215      & 8 2 3 4 2 5 2     \\
8       & 3.930471325 & 3.930478        & 9 2 3 3 4 2 5 2  \\
6       & 3.937517    & 3.937596      &  8 2 3 3 4 2        \\
8       & 3.9442122   & 3.944217375  & 9 2 3 3 5 2 4 2  \\
7       & 3.951028    & 3.951046     & 8 2 3 3 5 2 3     \\
4       & 3.960102    & 3.960768                  & 6 2 3 3              \\
8       & 3.960768675 & 3.96109855          & 8 2 3 3 6 2 3 3  \\
7       & 3.968975    & 3.968984      & 8 2 3 3 5 2 3     \\
8       & 3.973723775 & 3.973725725  & 9 2 3 3 5 2 4 2  \\
6       & 3.977761    & 3.977784       & 8 2 3 3 4 2        \\
8       & 3.98140865  & 3.9814099    & 9 2 3 3 4 2 5 2  \\
7       & 3.984747    & 3.98475       & 9 2 3 3 3 4 2     \\
8       & 3.9877453   & 3.987746125  & 9 2 3 3 3 5 2 3  \\
5       & 3.990258    & 3.990296      & 7 2 3 3 3           \\
8       & 3.9925194   & 3.99251987  & 9 2 3 3 3 5 2 3  \\
7       & 3.994538    & 3.9945385     & 9 2 3 3 3 4 2     \\
8       & 3.99621955  & 3.99621976425 & 10 2 3 3 3 3 4 2 \\
6       & 3.997583    & 3.9975846     & 8 2 3 3 3 3        \\
8       & 3.998641656 & 3.9986416967  & 10 2 3 3 3 3 4 2 \\
7       & 3.999397025 & 3.99939715     & 9 2 3 3 3 3 3     \\
8       & 3.99984936  & 3.999849365   & 10 2 3 3 3 3 3 3\\
\hline
\end{tabular}
\end{table}

\section{\label{sec42:level2}Elementary block frequencies}

Knowing that any periodic time series and the chaotic series at the onset of chaos of any of the period-doubling cascades have a well determined visibility pattern, how can we classify a given time series into one of these regimes bby simply analysing its visibility pattern? A possibility is to compute the frequency of the elementary blocks in the time series and to compare it with the elementary blocks of theoretical time series as shown in previous sections.

It is not difficult to prove that the frequency of any elementary block of size $m-1$ of the cascade of basic period $m \geq 2$ in the visibility pattern of a times series of period $T=m\, 2^{n-1}$ is given by:
\begin{equation}\label{nu}
\nu_n^m = \frac{2^{n-1}}{m \, 2^{n-1} -1}
\end{equation}
yielding a way of calculating the frequency of this elementary block at the onset of chaos of a cascade of basic period $m$:
\[
\lim_{n\to\infty} \nu_n^m = \frac{1}{m}
\]

\begin{figure}
\centering
\includegraphics[width=1.0\columnwidth]{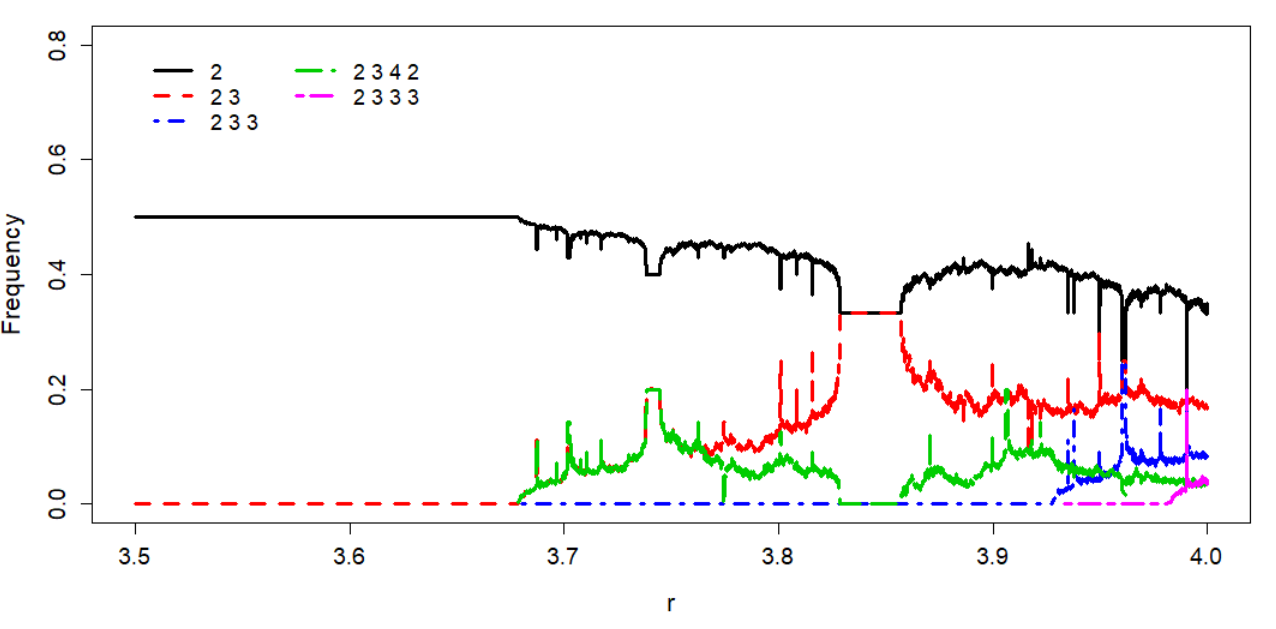}
\caption{Frequencies of the elementary block of the lowest period cascades as a function of $r$ for the logisitic map. The frequency of each block is calculated using time series of 5000 points with an $r$-increment of $10^{-4}$. For the periodic windows, the frequency of the corresponding elementary blocks coincides with the value obtained with expression \ref{nu}. Nonetheless, as it can be seen, the elementary blocks also appears out of their specific windows with a certain probability. It is important to mention that the frequency of the elementary blocks for the $m=2$ and $m=3$ cascades clearly detect the Misiurewicz point, in accordance with the theorem that assures that no cycles of period 3 exist below this point in unimodal maps. Besides, this plot uncovers other unknown distribution of the elementary blocks as, for example, the null presence of $C_5^1$  in the 3-period window, where the elmentary block $C_3$ achieves its largest frequency.}
\label{gen}
\end{figure}

Besides this analytical calculation of the frequencies of elementary blocks in the respective periodic windows, it is also possible to estimate the frequencies of the elementary blocks of any given time series and compare them with the theoretical predictions. For example, figure \ref{gen} shows the frequencies of the elementary blocks of the lowest periodic cascades as a function of $r$ for the logistic map. In the chaotic regions, these distributions cannot be deduced directly from the period doubling process except for the respective accumulation points at the onset of chaos. 

It is interesting to point out the change in the frequencies of the elementary blocks in the Misiurewicz point. The frequencies of all elementary blocks different from $\{2\}$, the elementary block of the Feigenbaum cascade, are null. The same occurs for the elementary blocks of larger period windows (data not shown). Also note that the complementarity in the frequencies of elementary blocks. For instance, the prevalence of the elementary block $\{2 \, 3\}$ in the 3-period doubling cascade is clearer when compared with the frequencies of the other elementary blocks. Finally,  the frequency of the blocks enables to detect the corresponding period window, as shown in figure \ref{gen} with respect to the period 4 elementary block $\{2 \, 3 \, 3 \}$, that presents a well defined peak for $r \approx 3.96$.

The distribution of elementary blocks as a function of $r$ for the logistic map complements the already known result relating to the distribution of visibilities of chaotic times series \cite{Lacasa2}: it is shown that the distribution is exponential with an exponent that depends on $r$, but that it is always lower than $\ln{\frac{3}{2}}$. In particular, 
in respect to the onset of chaos of the Feigenbaum cascade: $r=3.67$, this distribution is $f_{3.67} =0.3825\, e^{-0.312\,v}$ with an $R^2=0.9704$, whereas for the onset of chaos of the 3-period window: $f_{3.8495} (v) = 0.972 \, e^{-0.387 \,v}$ with $R^2 = 0.9859$ (in both cases, using a time series of size $N=2\,10^4 $).

%\begin{widetext}

\begin{table}[tbp]
\caption{Frequency of visibility for different $r$-values: (i) $r=3.67$ and $r = 3.8495$. The averages
 are obtained from time series with 20000 points}  
\label{tab9}
 \small
\begin{tabular}{lll}
\hline  
\textbf{v}  &  \textbf{$f_{3.67}$} & \textbf{$f_{3.8495}$} \\
\hline
2 &	0.5 & 0.3333  \\
3 & 0.0 & 0.3334 \\
4 &	0.18215  & 0.0 \\
5 &  0.1358 & 0.1666\\
6 &	0.0541 & 0.0\\
7 &  0.03465 &  0.0825 \\
8 &	0.02165 & 0.0 \\
9 &  0.01815 & 0.04165  \\
10 &   0.0136 &  0.0 \\
11 & 0.011 & 0.015 \\
12 & 0.0705 & 0.01165 \\
13 & 0.0055 & 0.00375\\
14 & 0.0048 & 0.0031 \\
15 & 0.0029 & 0.0017 \\
\hline  
\end{tabular}
\end{table}
%\end{widetext}

Note that some visibilities are not present in the time series. This is a consequence of the generation process of the visibility patterns in each of the periodic windows. 

\section{\label{sec42:level2}Concluding remarks}

Time series can be seen as a one-dimensional discrete contour in the plane. They can also be studied from a geometrical point of view by measuring their visibility properties. In particular, by applying a horizontal visibility mapping, we can associate each point of the series $(t,x(t))$ to a number that corresponds to its basin of visibility, i.e. the number of points that see it. The main characteristics of this mapping for different time series have been proven in previous papers \cite{Luque1, Lacasa2, Luque2, Lacasa18}.

In this paper, we have focused the study on the visibility patterns of time series that are solutions of unimodal maps. These visibility patterns are formed by the visibility of each point of the series.  In particular, we have shown that: (i) Temporal patterns of visibility exist in the time series derived from unimodal maps. (ii) there is a universal procedure that generates the visibility patterns for all time series in the period-doubling cascades of the bifurcation diagram of unimodal maps. 
(iii) there are elementary blocks for each periodic window of the bifurcation diagram, (iv) from these elementary blocks, the visibility patterns of each period within the period-doubling cascades are formed by applying the universal procedure, (v) at the onset of chaos of each period-doubling cascade of any period, there is a visibility pattern of the chaotic time series,  containing all infinite previous patterns and (vi) the determination of all elementary blocks for each period doubling cascade for greater periods is always accessible and is only limited by computational costs.

The visibility patterns we have described in this paper are a kind of symbolic representation of the time series. Symbolic dynamics was proposed as an alternative tool to study the properties of dynamical systems \cite{Hao}. The symbolic dynamics of a chaotic dynamical system is, in most cases, intimately related to its geometrical structure.  In general, two maps that share the same symbolic properties are topologically conjugate, which means that their bifurcation diagram shows the same bifurcations in exactly the same order (\cite{Gilmore}). More recently, the symbolic dynamics induced by the visibility algorithm have been further studied in the logistic map to characterize chaos  (\cite{Lacasa18}). In this context, sequential motifs have been defined in the visibility graphs of time series (\cite{Iacovacci1,Iacovacci2}). These motifs are subgraphs formed by consecutive nodes in a Hamiltonian path and provide a kind of dynamic symbolization of the time series. However, in contrast to the visibility patterns described in this paper, which are temporaly ordered, these motifs are defined in the visibility graph and, therefore, they do not take into account the time sequence.

The visibility patterns can be applied  to characterize time series of unkown origins, in particular, thoses that are either chaotic or random. As we have shown, chaotic time series that are generated by unimodal maps have visibility patterns that yield well determined number distributions (see the tables in the main text). This can be used to distinguish these deterministic series from other kind of time series, such as random or brownian. 
Besides, the determination of the visibility patterns for a chaotic series can classify the series whithin the corresponding band of the bifurcation diagram. For instance, any time series with a visibility pattern having traces of visibility 3 must belong to the principal band, beyond the Misiurewicz point (\cite{Misiurewicz}). In the logistic map, the Misiurewicz point is reached at $r=3.67857351042832226...$, just the point at which the frequency of 3-visibility points in the time series becomes larger than zero (see Fig. \ref{gen})

The visibility patterns that have been uncovered in this paper determine the procedure to build partial visibility curves (\cite{Nuno20}). Recall that a partial visibility curve is formed of points $(\nu,v(\nu))$, with $v(\nu)$ being the maximum numbers of points of the time series that are seen from the set $\nu$ (see Fig. 4 of \cite{Nuno20} to visualize the process of point selection.) For each $\nu$, knowing the visibility pattern provides a method to choose  the points of the time series with largest visibility. However, the complete generation of the visibility curve cannot be carried out exclusively with the visibility pattern because of redundancies, i.e. non-null intersections of the visibility basins of the points of the visibility set, that appear when $\nu$ reaches certain values. 

The visibility patterns, as other ordered structures that are present in the dynamics of unimodal maps, show that, indeed, there is certain order within chaos. Among the characteristic properties of chaos are the sensitivity to initial conditions and the unpredictability of the trajectories. However, as we have shown in this work, there is a regularity with regards to visibility: (i) trajectories with different initial conditions share a similar visibility pattern and (ii), once the elementary block is detected as a function of the parameter of the unimodal map,  the visibility pattern of the whole period doubling cascade can be generated.  Knowing the visibility patterns provide important asymptotic properties of time series.

\section*{Acknowledgements}

\bibliographystyle{elsarticle-num}

\bibliography{\jobname}

\begin{thebibliography}{11}

\bibitem{Yorke} Tien-Yien Li and  Jame A. Yorke {\it Period three implies chaos},  The American Mathematical Monthly  Vol. 82, No. 10,  985-992 (1975)

\bibitem{Metropoli} Metropolis, N., Stein, M.L., and Stein, P.R. On finite limit sets for transformations
of the unit interval. J. Comb. Theory A, 15(1), 25–44. (1973) 

\bibitem{Shark} Sharkovsky, A.N. Coexistence of cycles of a continuous transofrmation of a line into itself, Ukrains’kii Mathematical Zhurnal, 16(1964), pp. 61–71
% Sharkoskii ordering. http://www.scholarpedia.org/article/Sharkovski

\bibitem{Gleick} Gleick, J. Chaos: Making a New Science, Viking, Penguin (1987) 


\bibitem{Strogatz} Steven H. Strogatz. {\it Nonlinear Dynamics and Chaos With Applications to Physics, Biology, Chemistry, and Engineering}. Westview Press-CRC Press (2018)

 \bibitem{Peitgen} Heinz-Otto Peitgen, Hartmut J\"urgens, Dietmar Saupe. {\it Chaos and Fractals New Frontiers of Science}. Second Edition-Springer (2004)

\bibitem{Schuster} Heinz Georg Schuster Wolfram Jus. {\it Deterministic Chaos: An Introduction} (Fourth, Revised and Enlarged Edition) WILEY-VCH Verlag (2005)


\bibitem{Gilmore}  Gilmore, R. and M. Lefranc. The Topology of Chaos: Alice in Stretch and Squeezeland, Wiley-VCH Verlag GmbH, 2nd Edition (2011)

\bibitem{Lacasa1} Lacasa, L., B. Luque, F. Ballesteros, J. Luque and J. C. Nu\~no, {\it From time series to  complex networks: the visibility graph }, Proc. Natl. Acad. Sci. USA 105 (2008) 4973-4975

\bibitem{Luque1} Luque B, Lacasa L, Ballesteros FJ, Robledo A. {\it Feigenbaum Graphs: A Complex Network Perspective of Chaos}. PLoS ONE 6(9): e22411. doi:10.1371/journal.pone.0022411 (2011) 

\bibitem{Lacasa2} Lacasa, L. and R. Toral. {\it Description of stochastic and chaotic series using visibility graphs}, Phys. Rev. E 82, 036120  (2010)


\bibitem{Luque2} Luque, B., L. Lacasa and A. Robledo. {\it Feigenbaum graphs at the onset of chaos}. Phys.Lett. A Vol. 376, Issues 47-48,  3625-3629 (2012)

\bibitem{Luque3} B. Luque, L. Lacasa, F. Ballesteros, and J. Luque {\it Horizontal visibility graphs: Exact results for random time series}, Phys. Rev. E 80, 046103 (2009)

\bibitem{Cady} Field Cady. {\it The data science handbook}, John Wiley \& Sons, Inc. (2017)

\bibitem{May} Robert M. May. {\it Simple mathematical models with very complicated dynamics}, Nature, 261, 459-467 (1976)

\bibitem{Feigenbaum} Feigenbaum M. J. {\it Quantitative universality for a class of nonlinear
transformations.}, J Stat Phys 19, 25 (1978)

\bibitem{Livadiotis} George Livadiotis. {\it High Density Nodes in the Chaotic Region of 1D
Discrete Maps}. Entropy 2018, 20, 24; doi:10.3390/e20010024

\bibitem{CRAN} R Core Team. {\it R: A language and environment for statistical computing}. R Foundation for Statistical
  Computing, Vienna, Austria. (2018) https://www.R-project.org/.

%\bibitem{Karamanos} K. Karamanos. {\it From Symbolic Dynamics to a Digital Approach Chaos}, M. Planat (Ed.): LNP 550, pp. 357-371, Springer-Verlag (2000)

\bibitem{Nuno20}  Nuño, J. C. and F. J. Muñoz. {\it The partial visibility curve of the Feigenbaum cascade to chaos}. Chaos, Solitons and Fractals 131 (2020) 109537
https://doi.org/10.1016/j.chaos.2019.109537

\bibitem{Galias} Zbigniew Galias and Bartłomiej Garda. {\it Detection of all low-period windows for the logistic map}. 978-1-4799-8391-9/15/ IEEE (2015)

\bibitem{Tucker} W. Tucker and D. Wilczak. {\it A rigorous lower bound for the stability regions of the quadratic map}, 
Physica D, vol. 238, no. 18, pp. 1923–1936, (2009)


\bibitem{Misiurewicz} Misiurewicz, M. {\it Absolutely continuous measures for certain maps of an interval}, Publications mathématiques de l’I.H.É.S., tome 53, p. 17-51 (1981)

\bibitem{Hao} Hao, Bai-Lin and Wei-Mou Zheng. {\it Applied symbolic dynamics and chaos}, Wolrd Scientifics Publishing (1998)

\bibitem{Lacasa18} Lacasa, L. and W. Just. {\it Visibility graphs and symbolic dynamics}. Physica D374–375, 35–44 (2018)


\bibitem{Iacovacci1} Iacovacci, J. and L. Lacasa. {\it Sequential visibility-graph motifs}. Physical Review E 93, 042309 (2016)


\bibitem{Iacovacci2} Iacovacci, J. and L. Lacasa. {\it Sequential motif profile of natural visibility graphs}. Physical Review E 94, 052309 (2016)

\bibitem{LosAlamos} https://library.lanl.gov/cgi-bin/getfile?00416636.pdf

\bibitem{Zou} Zou Y., R.V. Donner, N. Marwan, J. F. Donges  and J. Kurths. {\it Complex network approaches to nonlinear time series analysis}, Physics Reports,Volume 787, 21, Pages 1-97 (2019) 

\end{thebibliography}

\end{document}